# Measurement comparison among Time-Domain, FTIR and VNA-based spectrometers in the THz frequency range

L. Oberto[1], M. Bisi[1], A. Kazemipour[2], A. Steiger[2], T. Kleine-Ostmann[2], and T. Schrader[2]

[1]Istituto Nazionale di Ricerca Metrologica, strada delle Cacce 91, 10135, Torino, Italy,
l.oberto@inrim.it, m.bisi@inrim.it

[2]Physikalisch-Technische Bundesanstalt (PTB), Bundesallee 100, 38116 Braunschweig, Germany,
thomas.kleine-ostmann@ptb.de, andreas.steiger@ptb.de, alireza.kazemipour@ptb.de

*Abstract* - In this paper we present the outcome of the first international comparison in the terahertz frequency range among three different kinds of spectrometers. A Fourier-Transform infrared spectrometer, a Vector Network Analyzer and a Time-Domain Spectrometer have been employed for measuring the complex refractive index of three travelling standards made of selected dielectric materials in order to offer a wide enough range of parameters to be measured. The three spectrometers have been compared in terms of measurement capability and uncertainty.

## 1. Introduction

Terahertz (THz) radiation can pass through dielectric materials and is reflected back by conducting materials. Unlike X-rays, it is non-ionizing and allows for a variety of non-destructive tests giving access to significant material properties. Its potential for imaging is very high compared to microwaves, which offer a lower spatial resolution. On the other hand, it is more and more absorbed by water with increasing frequency [1]. Due to these unique properties, THz radiation has proved to be a valuable non-contact tool to gain insight into material properties [2].

Recently, THz technology has become more mature and ready-to-use: THz Time-Domain Spectrometers (TDS) are now commercially available; Fourier Transform Infrared (FTIR) spectrometers can also be adapted for measurements in the THz frequency range typically defined from 100 GHz to 10 THz; Vector Network Analyzers (VNA) have been utilized for THz dispersive spectroscopy, too [3, 4]. However, important properties and parameters such as the measurement uncertainty remain difficult to evaluate. Moreover, traceability of the main physical quantities to the international system of units (SI) is hardly available. The lack of measurement capabilities and methodologies for the main system parameters hinders the application of these technologies. Furthermore, missing metrological capabilities hamper the use of THz systems as measurement tools in fields such as security where reliability is crucial [5].

Because of the reasons mentioned above, instrument sellers do not quote the measurement uncertainty of their instruments, and measurements on the same material performed in different labs and with different techniques may not agree.

Given the availability of three different types of instruments, having a significant overlap of their frequency bands, a measurement comparison has been undertaken among three labs in two National Metrology Institutes (Istituto Nazionale di Ricerca Metrologica, INRIM, Torino, Italy, and Physikalisch-Technische Bundesanstalt, PTB, Braunschweig and Berlin, Germany). The aim of this work is to evaluate the performance of these spectrometers for measurements of material properties in the THz gap, in particular for what concerns measurement uncertainty. To the best of the authors' knowledge, no other comparison of the results obtained from the same material samples with different types of instruments has been performed apart from two comparisons between a TDS and a VNA [6] and [7], the latter giving not fully compatible results. A multi-lab intercomparison among TDS is currently ongoing [8], but final results are not available yet.

## 2. Measurands and travelling standards

The measurand chosen for this intercomparison was the complex refractive index of the materials under test. In particular, participants had to measure the refractive index and the extinction coefficient, *i.e.* the real and the imaginary part of the complex refractive index.

Travelling standards have been chosen in order to ensure wide enough ranges in terms of refractive index and extinction coefficient. Selected materials were fused quartz, pyrex and BK7. The travelling standards were optically-polished, thin disks having diameters of 100 mm (quartz) and 76 mm (pyrex and BK7). Their thicknesses were mechanically measured at INRIM. The results are summarized in Table 1.

| Material | Thickness / μm |
|----------|----------------|
| BK7      | 545 ± 1        |
| Pyrex    | 570 ± 1        |
| Quartz   | 549 ± 1        |

Table 1. Thickness of the travelling standards.

## 3. Employed instrumentation

A THz TDS has been used at INRIM. The spectrometer has been set up in the framework of a Joint Research Project of the European Metrology Research Program (EMRP), entitled Microwave and Terahertz Metrology for Homeland Security (JRP NEW07 THz Security). A picture of the system is shown in Figure 1.

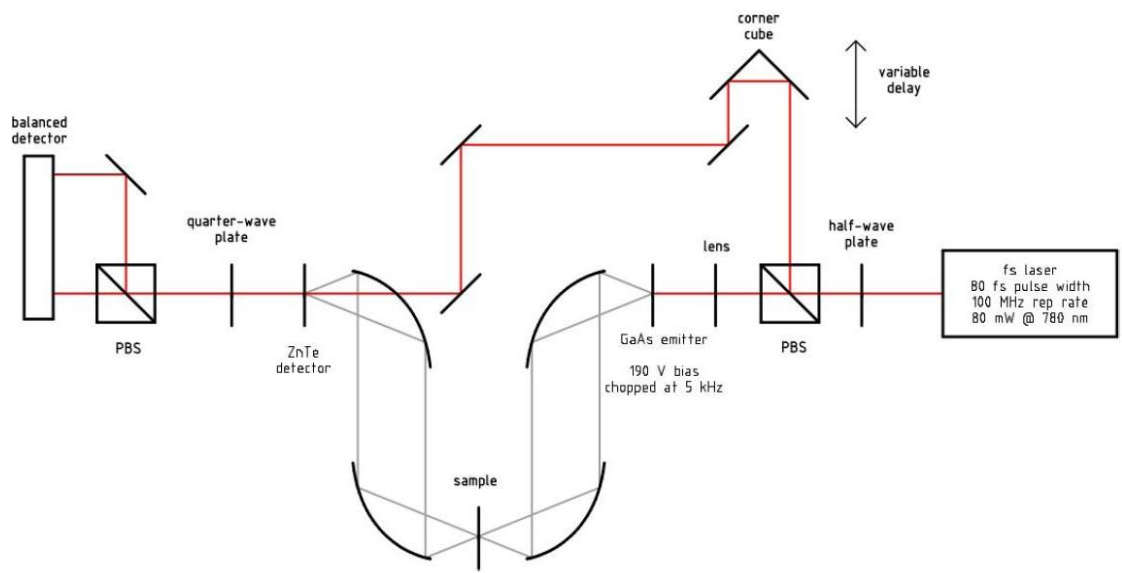

Figure 1. Schematic of the THz TDS at INRIM.

The TDS used a standard transmission configuration, with a photo-conductive antenna as THz emitter, a ZnTe crystal as detector, four off-axis parabolic mirrors to focus the THz beam onto the sample and a moving stage acting as delay line. It was capable of measuring from about 250 GHz to about 2 THz. The frequency resolution was about 29 GHz, a value set by the delay line path length, which was limited in order to perform fast measurements. A higher resolution was not needed because no absorption lines were expected. The delay line was characterized by means of a laser interferometer. The contribution to the SNR of the TDS due to the delay line position uncertainty has been addressed in [9]. The maximum bandwidth of our TDS, as set by the delay line position uncertainty, was found to be slighty below 3 THz. However the observed bandwidth was further limited by the emitter we used.

In order to reduce absorbing water vapor the TDS was placed in an almost-sealed box. Dehumidification was performed in a cost-effective and easy way by circulating the air in a closed loop by means of the orange pipe visible in Figure 1, and condensing the water vapor at a cold point inserted in the pipe. After about 1 hour, dry air was inserted in order to further reduce humidity. After less than 2 hours, a relative humidity of 5% was achieved. Humidity was constantly monitored during measurements by means of an hygrometer placed inside the box, close to the measurement point.

It is worth noting that the samples were placed in the system focal plane. It is known that, by positioning the samples in this way, a correction may be necessary to take into account the effect of the Gouy shift [10]. Anyway, after an *a priori* evaluation on the base of the estimated beam waist, we found this contribution to be negligible in our case with respect to the measurement uncertainty at low frequencies (less than 0.5%) while it vanishes at high frequency.

The measurement principle relies on the Fourier transformation of pulsed THz radiation. To extract the complex refractive index from the measured data, the ratio of the THz pulse spectra taken with ($E_{sam}(\omega)$) and without ($E_{ref}(\omega)$) the sample in the THz beam path was used to calculate the transfer function $H(\omega)$:

$$H(\omega) = \frac{E_{sam}(\omega)}{E_{ref}(\omega)}. \qquad (1)$$

The refractive index $n(\omega)$ and the extinction coefficient $k(\omega)$ were calculated from the following equations [9]:

$$\begin{aligned} n(\omega) &= n_0 - \frac{c}{\omega l} \angle H(\omega) \\ k(\omega) &= \frac{c}{\omega L} \left\{ \ln\left[\frac{4 n(\omega) n_0}{(n(\omega)+n_0)^2}\right] - \ln|H(\omega)| \right\} \end{aligned}, \qquad (2)$$

where $n_0$ is the refractive index of air, $c$ is the speed of the light and $L$ is the sample thickness. The contributions due to the parameters appearing in Eq. (2) were taken into account in the uncertainty analysis, along with the contribution of other parameters such as Fabry-Pérot reflections inside the sample (ignored by the model used for the optical parameter extraction) and the incidence angle on the sample of the THz beam [11]. It is worth noting that method in [11] ignores the reflections inside the sample in writing the transfer function, therefore they must be explicitly considered as an uncertainty source (see [11], equations 25 and 26). Moreover, multiple measurements have been performed. Their average and standard deviation have been taken into accout according to [11]. The stability of the system during the time scale of the measurement has also been checked by tacking a reference measurement before and another after the actual measurements, and by looking for any change. Anyway, no relevant effects have been noted.

For both refractive index and extinction coefficient, the highest contributions to the uncertainty come from the reflections inside the sample and from the thickness of the sample itself, while the contributions due to the approximation of the transfer function needed for obtaining (2) [11], to the sample alignment and to the air refractive index are less influential. An example of uncertainty budget is shown in Section 5.

Within the same EMRP project, PTB developed a VNA-based THz spectrometer, which is shown in Figure 2. Its frequency coverage is band-limited by the frequency converters used. It is capable of measuring in the range from 50 GHz to 500 GHz [12]. For this intercomparison it has been used in the frequency range 300-500 GHz with the frequency extension modules for the WR-3 and WR-2.2 waveguide bands setting a frequency resolution of approximately 55 MHz. Conventional rectangular horn antennas were mounted to the waveguide ports and two symmetrical parabolic mirrors were used to reduce the free-space transmission path-loss. Their distances to the antennas must be precisely determined in order to satisfy the plane-wave conditions on the material-under-test surface and to guarantee maximum energy passing through for a further improved sensitivity. The horn antennas have to be positioned so that their phase centers coincide with the focal points of the parabolic mirrors. The phase center positions can be estimated as a function of frequency, e.g., according to [13]. The frequency extension modules are positioned based on the distances obtained for the center frequency of the respective waveguide band. As shown in Fig. 2 a), the parabolic mirrors convert a diverging beam (assumed to originate from the phase center of the antenna) to a collimated beam at the sample position. The horn antennas are symmetrically put at the focal distance $f_0$ of parabolic mirrors which themselves are separated by a distance equal to $2f_0$ providing a plane wave at the middle of the two mirrors where the material-under-test is placed. The distance between the mirrors has to be $2f_0$ in order to obtain minimum phase distortion [14]. The system is calibrated at the flanges of the frequency extension modules as needed for material parameter extraction and tested with regard to stability as part of the uncertainty analysis (see Section 5).

For data extraction, a new concept is used [12]. It consists in virtually shifting the calibration plane from the VNA waveguide ports to the material-under-test surface. This represents a complete de-embedding technique for which the overall scattering effect of the propagation path from the actual VNA ports to the material surface is evaluated experimentally. The propagation path loss/diffraction is evaluated by measuring the S-parameters of the empty setup, which is supposed to be symmetrical. Then the same process is repeated with the material, in order to extract its parameters.

The permittivity and permeability of the material can be extracted by taking into account multiple-reflection effects in the material slab. Concerning the applied data-extraction method, both $S_{11}$ (which is assumed equal to $S_{22}$ due to symmetry) and $S_{21}$ (which is assumed equal to $S_{12}$ due to symmetry) are used if both permeability and permittivity are to be determined or uniquely the $S_{21}$ (or the mean of $S_{21}$ and $S_{12}$) is used for the transmission-only method when only the permittivity is needed.

The complex permittivity $\varepsilon_r = \varepsilon_r' - j\varepsilon_r''$ can be calculated from the slope $a$ of the phase of the transmission $T$ using

$$\varepsilon_r' = \left( \frac{a}{-j2\pi\sqrt{\varepsilon_0 \mu_0} L} \right)^2 \qquad (3)$$

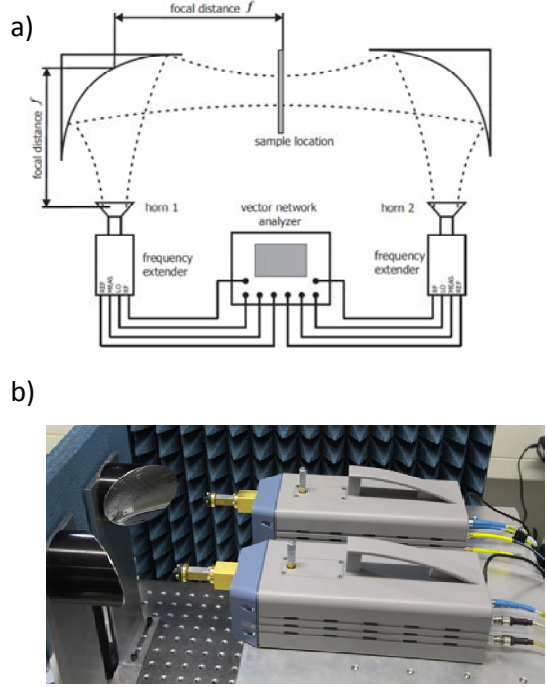

Figure 2. a) Schematic and b) picture of the VNA-based spectrometer at PTB.

and

$$\varepsilon_r{''} = -\frac{c_0 \ln|T|\sqrt{\varepsilon_r{'}}}{\pi f L}, \qquad (4)$$

where $L$ is the slab thickness, $f$ is the frequency and $\varepsilon_0$ and $\mu_0$ are the electric and magnetic field constants, respectively. Because of multiple reflection effects inside the material slab, the measured S-parameters are different from the actual transmission values $T$ and reflection values $\Gamma$ of the slab. Starting with $T = S_{21}$, a first estimate of $\varepsilon_r$ is obtained. This is used to get a better estimate of $T$ and $\Gamma$. A new intrinsic material parameter estimate of $\varepsilon_r$ is then determined from $T$ and $\Gamma$, and so on, whereas the result converges after 1 or 2 iterations. From the complex permittivity $\varepsilon_r$, the complex refractive index $\underline{n}$, consisting of real part $n$ and imaginary part $\kappa$, are determined with $\underline{n} = \sqrt{\varepsilon_r}$.

The errors relative to the factors $L$ and $a$ can affect the deduced permittivity. These effects can be observed as regular oscillations on the resulting curves especially for the imaginary part of a material slab which is not homogenous and has a non-uniform surface thickness. The oscillations appear as the removal of Fabry-Pérot resonances with the iterative method does not work very well if thickness varies or the sample is inhomogeneous. In many cases, the non-uniformity of the material surface and the thickness variation $\delta L$ are the most important contributions to the $\varepsilon_r{''}$ uncertainty. The uncertainty increases if the thickness (electrical length) of the sample gets smaller.

PTB also used a commercial FTIR spectrometer (Bruker Optics – Vertex 80v) adapted for measuring at THz frequencies. An optimized setup for the far infrared spectral region consisted of a high-pressure mercury discharge lamp as a powerful external THz radiation source. In contrast to its standard mounting directly at an external input port of the spectrometer, this high-power discharge lamp is connected via an off-axis parabolic mirror to a polyethylene (PE) entrance window which is transparent for the far infrared radiation but blocks the unwanted UV radiation of the discharge lamp. The external beam bath between the discharge lamp and PE entrance window is encased enabling a nitrogen purge to avoid water vapor absorption of the THz radiation. In addition, a pyroelectric THz detector at room temperature records the interferogram of the FTIR spectrometer. A room-temperature detector does not suffer from thermal background radiation of the sample at room temperature in contrast to cryogenic bolometers which are often used as sensitive detector for THz radiation [15]. Moreover, the good linear response of pyroelectric detectors prevents the occurrence of artifacts caused by nonlinear signal recording in the spectra after Fourier transformation. These advantages over bolometers at cryogenic temperatures are accompanied by a limited capability of measuring signals at very low powers caused by strong absorbing samples even if the high-pressure mercury discharge lamp is applied as a powerful external THz radiation source.

The samples are mounted on a motorized translation stage at an intermediate focus inside the compartment of the spectrometer. Vacuum inside the spectrometer avoids water vapor absorption similar to the nitrogen purge outside. Therefore, this instrument is capable of measuring in a broad spectral range down to approximately 200 GHz. The measurement principle is based on Fourier transformation of the recorded interferogram generated by the Michelson interferometer inside the FTIR spectrometer. The samples had to be slightly tilted off normal incidence in order to avoid spectral artifacts by specular reflection of the modulated THz radiation back to the interferometer. In order to get low noise spectra an average of 256 scans was recorded.

The frequency accuracy of the FTIR spectrometer with a relative uncertainty less than $2 \times 10^{-6}$ is assured by a red He-Ne laser of known absolute wavelength (632.816 nm) which is integrated into the beam path of the interferometer. The He-Ne interference pattern is used to linearize the movement of the mirror of the spectrometer for each interferogram. The spectral resolution is determined by the maximum travel of mirror movement which limits the resolution of this spectrometer to 6 GHz. The resolution was set to 15 GHz in this experiment because there are no sharp spectral features of the three samples in the spectral range. The power stability of the optimal setup with temperature stabilization of the whole instrument including the external high-pressure mercury discharge lamp is better than 1 % after a warm up time of one hour.

The transmission spectrum of the sample is calculated as the ratio of two measurements, with and without the sample in the optical path. As an example, the transmission spectrum of quartz is shown in Figure 3.

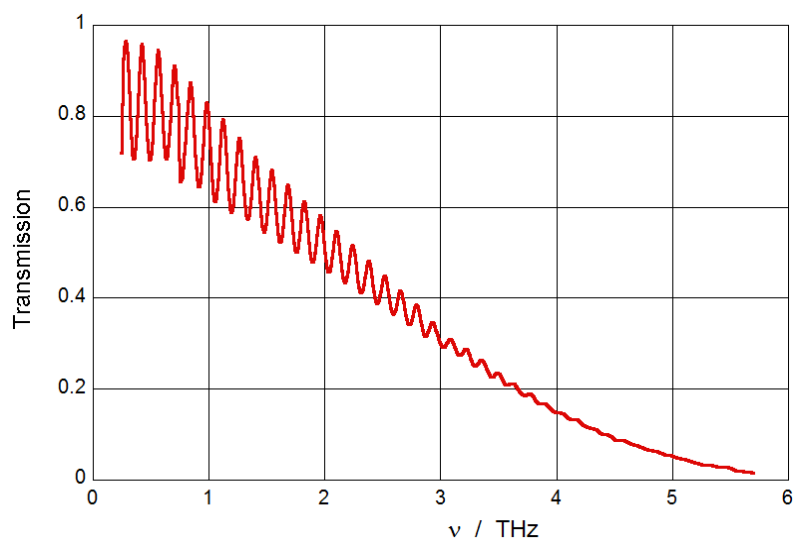

Figure 3. Transmission spectrum of quartz measured with the FTIR spectrometer.

The thin samples allowed the extraction of the refractive index with low uncertainty simply from the frequency position of the Fabry-Pérot fringes of the transmission spectra without applying a complex analytical model as described in [16]. The absolute interference order is easily determined because the free spectral range of a 0.5 mm thin sample is only about one half of the lower frequency limit of the recorded spectra. Just these frequency positions $\nu(m)$ of both the maxima (integer order: m) and the minima (half integer order: m+1/2) are used to calculate the frequency dependant refractive index $n(\nu)$ from the known thickness $L$, the speed of light $c$, and the tilt angle $\delta$ of the beam inside the sample:

$$n(\nu) = \frac{m \cdot c}{2L \cdot \cos(\delta) \cdot \nu(m)}. \quad (5)$$

For a given thickness $L$, the uncertainty of $n(\nu)$ is simply given by two terms: the uncertainty of maxima/minima frequency position determination in the recorded spectra and a minor contribution by the uncertainty due the cosine of a small tilt angle. The spectral transmission signal is used to extract the spectral absorption. First of all only the maximum values are selected in the transparent region with visible Fabry-Pérot fringes. Then an interpolating curve $T$ is calculated from these maxima. Outside this frequency interval, the transmission is < 8 %, i.e. no signal modulation due to interference by standing waves inside the sample is visible in the data. Therefore the interpolating curve $T$ is extended simply by the measured data for $T < 0.08$. The absorption is then calculated as $-\ln(T)$. Finally, the extinction coefficient $k$ is calculated from the thickness $L$ and the frequency ν:

$$k(\nu) = -\ln(T) \cdot \frac{c \cdot \cos(\delta)}{L \cdot 4\pi \cdot \nu}. \quad (6)$$

Besides the uncertainty of the absorption $\ln(T)$ there are relative uncertainty contributions to the uncertainty of the extinction coefficient $k$ similar to the relative uncertainty of $n(\nu)$ namely the relative uncertainty of the thickness $L$, the cosine of the small tilt angle $\cos(\delta)$ and the frequency ν.

To better clarify ranges and resolutions of the three instruments, these data are summarized in Table 2.

| Instrument | Freq. Range / THz | Resolution |
|---|---|---|
| TDS | 0.25-2.5 | 29 GHz |
| VNA | 0.3-0.5 | 55 MHz |
| FTIR | 0.2-6 | 15 GHz |

Table 2. Frequency ranges and resolutions of the three instruments used.

### 4. Measurement results

In this Section, the measurement results are compared in form of graphs. The comparison between TDS, FTIR and VNA spectrometer, in the lower part of the THz frequency range (i. e. below 500 GHz), has been performed on two samples only (quartz and BK7), as pyrex was not measured in the VNA spectrometer. The comparison between TDS and FTIR spectrometer has been made among measurements concerning all the travelling standards, but the FTIR measurement of the refractive index of BK7 has been made at the frequency of 400 GHz only, due to high material absorption at higher frequencies. Here and in the rest of the paper, reported uncertainties are always expanded with coverage factor $k = 2$ for a 95% confidence interval [17].

#### 4.1. Quartz

As seen from Figure 4, FTIR measurements span over the widest frequency range. The band covered by the VNA spectrometer is the smallest one while TDS is in between. TDS and FTIR spectrometer have roughly comparable refractive index uncertainty while that of the VNA spectrometer is larger. On the other hand, the TDS has the largest extinction coefficient uncertainty and the FTIR spectrometer clearly outperforms the other instruments. Anyway, all measurements on the quartz sample agree within the specified measurement uncertainty. They also in agreement with the literature result of the quartz refractive index: n = 1.95 ± 0.05 [18].

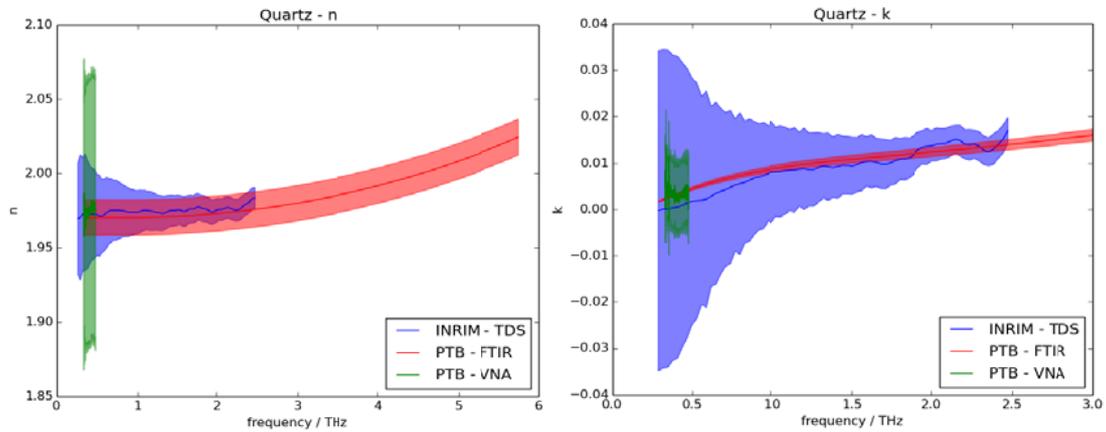

Figure 4. Comparison of measurement results for the quartz travelling standard.

### 4.2. BK7

In Figure 5, the measurement results for the BK7 travelling standard are shown. For this sample, the extinction coefficient is about ten times higher than that of the other two. Apparently, the INRIM TDS is the most suitable instrument for characterizing this kind of materials, as it is able to measure up to 1.1 THz with the lowest uncertainty. The FTIR spectrometer is severely limited by the material absorption. Despite the smaller uncertainty for the extinction coefficient in the low frequency range, the VNA-based spectrometer has the largest uncertainty for the refractive index, and is limited by its frequency coverage. As before, the comparison shows that all the measurements are well compatible within the specified measurement uncertainties. They are also compatible with the literature result of the BK7 refractive index: n = 2.5 ± 0.1 [18].

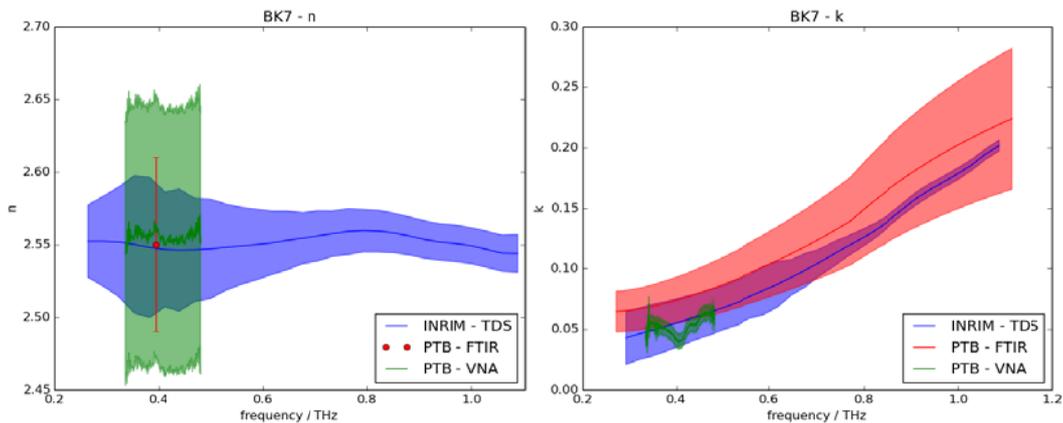

Figure 5. Comparison of measurement results for the BK7 travelling standard.

### 4.3. Pyrex

The pyrex travelling standard has been measured by PTB with the FTIR spectrometer only. Results are shown in Figure 6 and again, they are well compatible. As already pointed out, the FTIR spectrometer has a wider frequency coverage compared to the TDS. The refractive index uncertainty is similar for both systems at high frequencies at least, while the extinction coefficient uncertainty is significantly better for the FTIR spectrometer. The literature result of the pyrex refractive index is: n = 2.1 ± 0.1 [18]. This value is not compatible with the one of this work. Since we do not know the composition of the borosilicate glass used in [18], a possible explanation of this discrepancy is the change in refractive index that occurs when borosilicate glasses have different composition.

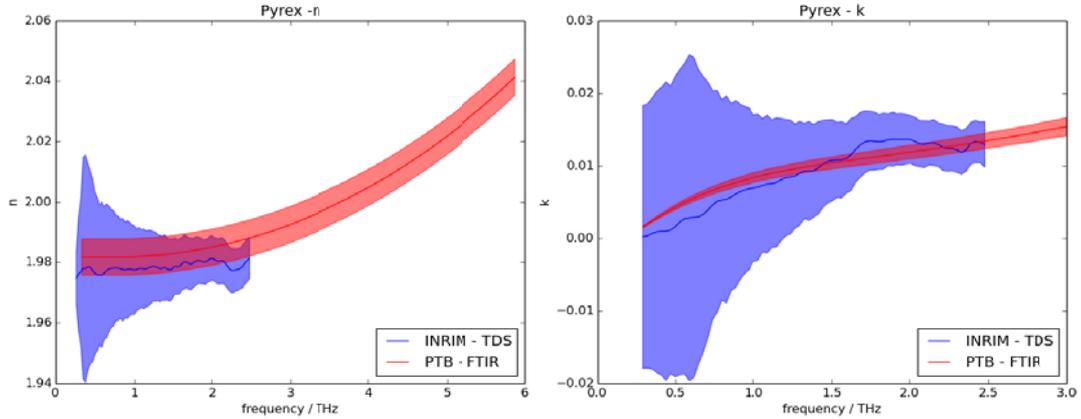

Figure 6. Comparison of measurement results for the pyrex travelling standard.

## 5. Discussion

As can be seen from the measurement results shown in Section 3, the measurement of the refractive index did not cause particular problems. The TDS showed better performances, in terms of measurement uncertainty, in measuring the refractive index compared to the VNA spectrometer in the band of overlapping and slightly better performance compared to the FTIR spectrometer.

The measurement of the extinction coefficient was more critical. Apart from FTIR, there is a difficulty in measuring the extremely low absorption of quartz and pyrex, almost transparent at lower frequencies *i.e.* below about 800 GHz. The VNA-based spectrometer performs slightly better than the TDS but the FTIR spectrometer is by far the best choice. However for BK7 with a ten times higher absorption, the TDS exhibits the smallest uncertainty above 500 GHz.

Although the FTIR spectrometer has the widest frequency coverage, this instrument is not capable to determine the BK7 refractive index from transmittance measurements in the complete frequency band due to its high absorption. Only a single point evaluation at 400 GHz could be performed. Moreover, the FTIR measurement of BK7 reveals a high uncertainty of the extinction coefficient due to the low transmitted signal caused by the strong absorption of the sample.

The VNA-based spectrometer turned out to be a good choice for measurements in the lower part of the THz frequency range and below, while FTIR spectrometers are more suitable for measurements of the extinction coefficient in the band of interest and beyond. THz Time-Domain spectrometers turned out to be the best choice for measuring the refractive index in the whole band from about 100 GHz to about 2.5 THz. At lower frequencies VNA-based spectrometers should be preferred while FTIR spectrometers are needed above 2.5 THz.

Tables 3, 4 and 5 show a comparison of the measurements for the three samples at the common frequency of 400 GHz.

| Instrument | Refractive index | Extinction coefficient |
|---|---|---|
| INRIM TDS | 1.98 ± 0.03 | -0.002 ± 0.026 |
| PTB VNA | 1.98 ± 0.09 | 0.003 ± 0.008 |
| PTB FTIR | 1.970 ± 0.006 | 0.0032 ± 0.0003 |

Table 3. Comparison of fused quartz measurements at 400 GHz.

| Instrument | Refractive index | Extinction coefficient |
|---|---|---|
| INRIM TDS | 2.55 ± 0.03 | 0.05 ± 0.01 |
| PTB VNA | 2.56 ± 0.09 | 0.042 ± 0.007 |
| PTB FTIR | 2.55 ± 0.06 | 0.07 ± 0.02 |

Table 4. Comparison of BK7 measurements at 400 GHz.

| Instrument | Refractive index | Extinction coefficient |
|---|---|---|
| INRIM TDS | 1.98 ± 0.02 | -0.005 ± 0.021 |
| PTB VNA | --- | --- |
| PTB FTIR | 1.982 ± 0.006 | 0.0031 ± 0.0002 |

Table 5. Comparison of pyrex measurements at 400 GHz. This standard was not measured with the VNA-based spectrometer.

As an example, detailed uncertainty budgets for the three instruments are reported in tables 6, 7 and 8. Data refer to the fused quartz measurements at 400 GHz reported in Table 3.

| Contribution | $n$ | $k$ | Type |
|---|---|---|---|
| Sample thickness | 1.8E-03 | 5.21E-04 | A |
| Sample measurement noise and repeatability | 3.6E-04 | 1.15E-04 | A |
| Reference measurement noise and repeatability | 6.1E-05 | 3.01E-04 | A |
| Sample alignment | 1.5E-04 | 4.95E-06 | B |
| Transfer Function approximation | 1.1E-05 | 2.17E-06 | B |
| Fabry-Perot internal reflections | 2.5E-02 | 2.45E-02 | B |
| Air refractive index | 2.7E-04 | 2.82E-05 | B |

Table 6. INRIM TDS uncertainty contributions of the quartz measurements at 400 GHz. Contributions have been combined as per [11] eq. (30).

| Contribution | $n$ | $k$ | Type |
|---|---|---|---|
| Transmission amplitude (VNA) | 0.001 | 0.003 | A |
| Transmission phase (VNA) | 0.025 | 0.001 | A |
| Sample thickness | 0.086 | 0.001 | A |
| Data extraction | 0.006 | 0.007 | B |

Table 7. PTB VNA uncertainty contributions of the quartz measurements at 400 GHz.

| Contribution | $n$ | $k$ | Type |
|---|---|---|---|
| sample thickness | 1.8 ‰ | 1.8 ‰ | B |
| alignment ($\cos(\delta)$) | 1.2 ‰ | 1.2 ‰ | B |
| frequency determination | 2.5 ‰ | 2.5 ‰ | A |
| data extraction ($\ln(T)$) | − | 8 ‰ | B |
| total uncertainty | 3.3 ‰ | 9 ‰ | |

Table 8. PTB FTIR uncertainty contributions of the quartz results at 400 GHz.

## 6. Conclusions

The first bilateral measurement comparison on material properties in the THz frequency range has been conducted between INRIM and PTB. Travelling standards have been chosen in order to ensure wide enough ranges in terms of refractive index and extinction coefficient. Three different kinds of spectrometers *i.e.* Time-Domain, Fourier Transform Infrared and Vector Network Analyzer-based systems have been involved. These spectrometers employ different spectroscopic methods, namely Fourier transformation of pulsed radiation, Fourier transformation of broadband radiation and dispersive spectroscopy.

Results show good compatibility among the three measurement systems. The comparison between the performances of the different instruments involved frequency coverage and measurement uncertainty. It highlights the kind of instrument being best suited for every material measured.

## Acknowledgements

The INRIM spectrometer and the related data analysis software have been completed with the valuable help of S. Lippert and D. Jahn of the Department of Physics and Materials Sciences Centre at Philipps-Universität Marburg, Germany. This work has been conducted within the EMRP Joint Research Project Microwave and terahertz metrology for homeland security (JRP NEW07 THz Security). The EMRP is jointly funded by the EMRP participating countries within EURAMET and the European Union.

## References

[1] K. Sakai (Ed.), Terahertz Optoelectronics, Springer, Berlin, 2005.
[2] G. W. Chantry, Submillimetre Spectroscopy, Academic Press, 1971.
[3] B. Yang, X.e Wang, Y. Zhang, and R. S. Donnan, J Applied Physics **109**, 033509 (2011).
[4] W. Sun, B. Yang, X. Wang, Y. Zhang, and R. Donnan, Optics Letters **38**, 5438–5441 (2013).
[5] Publishable JRP Summary Report for JRP NEW07 THz Security Microwave and terahertz metrology for homeland security, available online: http://www.euramet.org/Media/docs/EMRP/JRP/JRP_Summaries_2011/New_Tech_JRPs/NEW07_Publishable_JRP_Summary.pdf
[6] T. Tosaka, K. Fujii, K. Fukunaga, A. Kasamatsu, IEEE Trans. THz Sci. Tech. **5**, 102–109 (2015).
[7] M. Naftaly, N. Shoaib, D. Stokes, N. Ridler, A comparison method for THz measurements using VNA and TDS, J Infrared Millim. Terahertz Waves **37**, 691–702 (2016).
[8] M. Naftaly, J. Molloy, A multi-lab intercomparison study of THz time-domain spectrometers, 40th International Conference on Infrared, Millimeter, and Terahertz Waves (IRMMW-THz 2015), Hong Kong, China, August 2015.
[9] D. Jahn, S. Lippert, M. Bisi, L. Oberto, J.C. Balzer, M. Koch, J Infrared Millim. Terahertz Waves **37**, 605–613 (2016).
[10] P. Kužel, H. Němec, F. Kadlec, C. Kadlec, Optics Express **18**, 15338–15348 (2010).
[11] W. Withayachumnankul, B. M. Fischer, H. Lin, D. Abbot, J. Opt. Soc. Am. B **25**, 1059–1072 (2008).
[12] A. Kazemipour, M. Hudlicka, S.K. Yee, M. Salhi, D. Allal, T. Kleine-Ostmann, T. Schrader, IEEE Trans. Instr. Meas. **64**, 1438–1445 (2015).
[13] E. Muehldorf, The phase center of horn antennas, IEEE Transactions on Antennas and Propagation **18**, 753-760 (1970).
[14] A. Gonzalez et al., A Horn-to-Horn Power Transmission System at Terahertz Frequencies, IEEE Transaction on Terahertz Science and Technology **1**, 416-424 (2011)
[15] M. Kehrt, C. Monte, AMA Conferences 2013 – IRS$^2$, 121-123 (2013), doi 10.5162/irs2013/iP7
[16] E. N. Grossman, D. G. McDonald, Opt. Eng. **34**, 1289-1295 (1995), doi:10.1117/12.199887
[17] BIPM, IEC, IFCC, ISO, IUPAC, IUPAP and OIML *Evaluation of Measurement Data Guide to the Expression of Uncertainty in Measurement JCGM 100:2008 (GUM 1995 with Minor Corrections)*, 1st edn (Sèvres: BIPM Joint Committee for Guides in Metrology), 2008
[18] M. Naftaly, R. E. Miles, Journal of Non-Crystalline Solids **351**, 3341–3346 (2005).